\documentclass[conference]{IEEEtran}
\IEEEoverridecommandlockouts
\usepackage{cite}
\usepackage{amsmath,amssymb,amsfonts}
\usepackage{algorithmic}
\usepackage{graphicx}
\usepackage{float} 
\usepackage{subfigure}
\usepackage{bm}
\usepackage{textcomp}
\usepackage{xcolor}
\usepackage{stmaryrd}
\usepackage{float}                  
\usepackage{hyperref}
\usepackage[ruled,linesnumbered]{algorithm2e}
\def\BibTeX{{\rm B\kern-.05em{\sc i\kern-.025em b}\kern-.08em
    T\kern-.1667em\lower.7ex\hbox{E}\kern-.125emX}}
\begin{document}

\title{Target Sensing With Off-grid Sparse Bayesian Learning for AFDM-ISAC System}

\author{\IEEEauthorblockN{Yirui Luo*, Yong Liang Guan*, Yao Ge$^\dagger$, and Chau Yuen*}
\IEEEauthorblockA{\textit{*School of Electrical and Electronic Engineering, Nanyang Technological University, Singapore } \\
\textit{$^\dagger$Continental-NTU Corporate Lab, Nanyang Technological University, Singapore } \\
yirui001@e.ntu.edu.sg, $\{$eylguan; yao.ge; chau.yuen$\}$@ntu.edu.sg.}
}

\maketitle

\begin{abstract}
The recently proposed multi-chirp waveform, affine frequency division multiplexing (AFDM), is regarded as a prospective candidate for integrated sensing and communication (ISAC) due to its robust performance in high-mobility scenarios and full diversity achievement in doubly dispersive channels. However, the insufficient Doppler resolution caused by limited transmission duration can reduce the accuracy of parameter estimation. In this paper, we propose a new off-grid target parameter estimation scheme to jointly estimate the range and velocity of the targets for AFDM-ISAC system, where the off-grid Doppler components are incorporated to enhance estimation accuracy. Specifically, we form the sensing model as an off-grid sparse signal recovery problem relying on the virtual delay and Doppler grids defined in the discrete affine Fourier (DAF) domain, where the off-grid components are regarded as hyper-parameters for estimation. We also employ the expectation-maximization (EM) technique via a sparse Bayesian learning (SBL) framework to update hyper-parameters iteratively. Simulation results indicate that our proposed off-grid algorithm outperforms existing algorithms in sensing performance and is highly robust to the AFDM-ISAC high-mobility scenario.
\end{abstract}

\section{Introduction}
A diverse array of emerging wireless services for the sixth-generation (6G) wireless systems, such as unmanned vehicles, V2X, and the industrial Internet of things (IIoT), require wireless networks to provide both target sensing and communication promptly \cite{Z.Zhang 6G wireless networks}. Simultaneously, communication and sensing systems are increasingly converging their frequency bands in response to the rising demand for fast communication and high-resolution sensing \cite{X. Fang}, which has led to competition for spectrum. Therefore, integrated sensing and communication (ISAC) aiming to combine both functions within a single system to improve spectrum and energy efficiency is considered promising to enable 6G wireless communications \cite{J. A. Zhang Survey}.

Most existing ISAC studies aim to integrate both communication and sensing by employing the same waveform and hardware. To achieve this, orthogonal frequency division multiplexing (OFDM) has been widely studied in ISAC due to the low detection complexity and robustness to target detection \cite{Y. Luo}. However, severe Doppler spreads caused by high mobility with high carrier frequency result in the loss of orthogonality between the subcarriers, which makes OFDM suffer from inter-carrier interference (ICI) \cite{Z. Wei OTFS}. Then, the communication performance is dramatically damaged. To address this issue, a chirp-based waveform called orthogonal chirp division multiplexing (OCDM) is proposed in \cite{OCDM}, whose performance is shown to outperform OFDM in time-dispersive channels due to the ability to exploit multipath diversity efficiently. However, it can not achieve full diversity in doubly dispersive channels \cite{AFDM conference}. To further improve the performance, orthogonal time frequency space (OTFS) modulation, a two-dimensional modulation technique, has also been studied for ISAC, e.g., \cite{W. Yuan}.  The OTFS modulates information symbols on the delay Doppler (DD) domain \cite{Yao} \cite{qinwen}, where different delays and Doppler shifts characterize the distinguishable paths of the wireless channel.

Recently, a new waveform based on the one-dimensional discrete affine Fourier transform (DAFT) named affine frequency division multiplexing (AFDM) has been introduced in \cite{AFDM TWC}. It has been proved that compared with OTFS, AFDM has comparable BER performance with less pilot overhead \cite{AFDM TWC,luo qu, yiwei} and lower modulation complexity \cite{K. R. R. Ranasinghe}. Besides, AFDM is regarded as a prospective candidate for ISAC due to the ability to ensure dependable communication in high-mobility scenarios and to distinguish multiple targets in the ISAC system. A challenging task for AFDM-based ISAC operation is the estimation of fractional Doppler with an energy leakage effect, which is caused by insufficient sampling of Doppler. To handle this estimation problem, the authors in \cite{Y. Ni ISAC} proposed a simple matched filter method for ISAC by using multiple AFDM symbols. However, the resolution for fractional Doppler estimation is limited by the number of AFDM symbols, which is undesirable due to the low-latency requirement. An approximate maximum likelihood (ML) algorithm is further proposed for performance improvement in \cite{A. Bemani ISAC}, but suffers from a high computational complexity in the exhausted grid search. Authors in \cite{K. R. R. Ranasinghe} and \cite{W. Benzine CS} exploited the channel sparsity of delay and Doppler in the DAF domain and have developed specific codebooks for sparse signal recovery. However, they can only estimate the on-grid parameters according to the corresponding codebook resolutions. The authors in \cite{K. R. R. Ranasinghe} emphasize that they could use a very dense sampling grid for the codebook to increase the accuracy of resolution, but it would also increase complexity as a cost. 

In this paper, inspired by the off-grid DOA estimation \cite{Z. Yang}, we take the off-grid Doppler component into consideration for sensing enhancement. Specifically, we formulate the sensing target parameter estimation problem in AFDM-ISAC system as a sparse signal recovery problem, which can be tackled by compressive sensing for joint range and velocity estimations. To address the resolution limit of the Doppler estimation, we propose a Bayesian framework for AFDM-ISAC, where the off-grid components of the Doppler shifts are regarded as hyper-parameters for estimation. We also employ an expectation–maximization (EM) algorithm in the sparse Bayesian learning (SBL) structure to update the hyper-parameters iteratively. Simulation results demonstrate that our proposed method achieves superior sensing performance compared to the state-of-the-art (SotA) algorithms in\cite{K. R. R. Ranasinghe} and \cite{W. Benzine CS} for the AFDM-ISAC system and shows strong robustness.

\section{AFDM ISAC System Model}
\begin{figure}
\centering
\includegraphics[width=8.5cm]{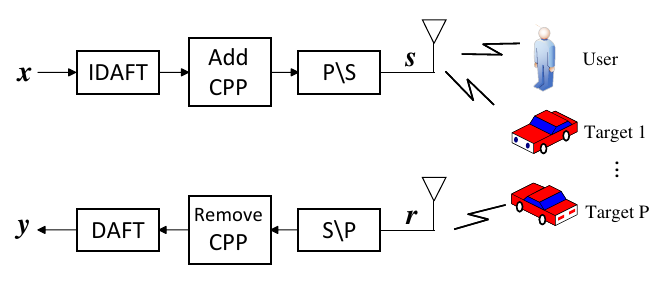}
\caption{AFDM-ISAC system model.}
\label{ISAC}
\end{figure}

In this paper, we consider an AFDM-ISAC system where a roadside unit (RSU) communicates with a specific user using AFDM communication signals while simultaneously receiving the echoes reflected by the targets in the surrounding environment. The RSU has a single ISAC transmit antenna and a co-located single ISAC receive antenna\footnotemark{}. The sketch of the system model for AFDM-ISAC is illustrated in Fig.~\ref{ISAC}. \footnotetext{Assuming there is sufficient isolation between the two antennas so that the echo of the sensing signal is not interfered by the transmitting signal\cite{W. Yuan}.}
\subsection{Transmitted AFDM Signal at the RSU}
Here, we consider a system with a bandwidth $BW=N \Delta f$, where $N$ denotes the number of chirp subcarriers and $\Delta f=1/T$ denotes the chirp subcarrier spacing. The duration of one AFDM symbol is $T$. Set $\mathbf{x}=\left[x[0],x[1],...,x[N-1]\right]^T$ to be the $N \times 1$ PSK/QAM symbol vector in the DAF domain. For transforming into the time domain, the $N$-point inverse DAFT (IDAFT) is applied, yielding

\begin{equation}
    s\left[ n \right] = \sum\limits_{m = 0}^{N - 1} {x\left[ m \right]{\phi _n}\left( m \right)},\text{  } n=0,\dots,N-1, \label{s[n]}
\end{equation}
with ${\phi _n}\left( m \right) = \frac{1}{{\sqrt N }}\exp \left[ {j2\pi \left( {{c_1}{n^2} + {c_2}{m^2} + nm/N} \right)} \right]$. $c_1$ and $c_2$ are AFDM parameters which is given in \cite{AFDM TWC}. In the matrix form,  \eqref{s[n]} can be expressed as ${\mathbf{s}} = {\mathbf{\Lambda }}_{{c_1}}^H{\mathbf{F}^H\mathbf{\Lambda }}_{{c_2}}^H{\mathbf{x}}$, where ${{\mathbf{\Lambda }}_{c_i}} = {\text{diag}}\left( {{e^{ - j2\pi {c_i}{n^2}}},n = 0,1,...,N - 1,\text{ }i=1,2} \right)$ and $\mathbf{F}$ is the normalized $N$-point DFT matrix.
Subsequently, to make the channel lie in the periodic domain, a chirp-periodic prefix (CPP) should be appended, which is defined as
\begin{equation}
    s[n] = s\left[ {N + n} \right]{{\text{e}}^{ - j2\pi {c_1}\left( {{N^2} + 2Nn} \right)}},{\text{     }}n =  - {N_{cpp}},..., - 1,
\end{equation}
where ${N_{cpp}}$ is CPP length, and it is an integer not less than the maximum channel normalized delay.

\subsection{Received Sensing Signal}
Assume there are $P$ targets around the RSU. The scattering coefficient, range, and velocity of the $i$-th target are denoted as $h_i$, $R_i$, and $V_i$, respectively. Then, the round-trip delay and Doppler shift of the $i$-th target are $\tau_{i}=2R_i/c$ and $f_{d,i}=2V_if_c/c$, where $f_c$ is the carrier frequency and $c$ is the speed of light. 

Then, the received sensing sample at time $n$ is denoted by
\begin{equation}
    r\left[ n \right] = \sum\limits_{i = 1}^P {{h_i}{{\text{e}}^{ - j2\pi {f_i}\left( {n - {\ell_i}} \right)}}} s\left[ {n - {\ell_i}} \right] + w\left[ n \right],
\end{equation}
with $n=0,1,...N+N_{cpp}-1$, $\ell_i=\tau_i/T_s$ and $f_i=f_{d,i}T_s$, where $T_s=T/N$ is the sampling interval. $w\left[n\right]$ is the $n$-th entry of the relative additive Gaussian noise vector $\mathbf{w}$. We define the normalized Doppler shift ${\nu _i} = {{f_{d,i}}} /\Delta f = {\alpha _i} + {a_i}$, where ${\alpha _i} \in \left[ { - {\alpha _{\max }},{\alpha _{\max }}} \right]$ is the integer normalized Doppler shift  and ${a_i} \in \left( { - \frac{1}{2}} \right.,\left. {\frac{1}{2}} \right]$ is the fractional normalized Doppler shift, respectively. The maximum delay satisfies $\ell_\text{max}\triangleq$ max$\left(\ell_{i}\right)$, which is smaller than $N_{cpp}$. 
 
 As proved in \cite{AFDM TWC}, $c_1$ is set to be
 \begin{equation}
     c_1=\frac{{2\left( {{\alpha _{\max }} + {k_v}} \right) + 1}}{{2N}}
 \end{equation}
to avoid different non-zero entries overlapping at the same position, and $k_v$ is defined in \cite{AFDM TWC}. Moreover,  
to achieve the optimal diversity order, the maximum delay and Doppler should satisfy \cite{AFDM TWC}
\begin{equation}
    2{\alpha_{\max }} + {\ell_{\max }} + 2{\alpha_{\max }}{\ell_{\max }} < N.
\end{equation}
 
After CPP removal, the received sensing signal in the time domain is converted back to the DAF domain via DAFT and can be expressed in matrix form as

\begin{equation}\label{yr}
    {\mathbf{y}} = \sum\limits_{i = 1}^P {{{\tilde h}_i}} {{\mathbf{\Lambda }}_{{c_2}}}{\mathbf{F}}{{\mathbf{\Lambda }}_{{c_1}}}{{\mathbf{\Gamma }}_{CP{P_i}}}{{\mathbf{\Delta }}_{{f_i}}}{{\mathbf{\Pi }}^{{\ell_i}}}{\mathbf{\Lambda }}_{{c_1}}^H{{\mathbf{F}}^H}{\mathbf{\Lambda }}_{{c_2}}^H{\mathbf{x}} + {\mathbf{\tilde w}},
\end{equation}
 where ${{\tilde h}_i} = {h_i}{e^{ - j2\pi {f_i}{\ell_i}}}$ and ${\mathbf{\tilde w}} = {{\mathbf{\Lambda }}_{{c_2}}}{\mathbf{F}}{{\mathbf{\Lambda }}_{{c_1}}}{\mathbf{w}}$ is the noise in the DAF domain. $\Delta_{f_i}$ and ${{\mathbf{\Pi }}^{{\ell_i}}}$ have the expression of $\Delta_{f_i}={\text{diag}}({e^{ - j2\pi {{\left( {0:{N} - 1} \right){f_i}}}}})$ and ${\mathbf{\Pi }}^{\ell_i}={{\mathbf{F}}^H}{\text{diag}}({e^{ - j2\pi {\left( {0:{N} - 1} \right) \ell_i}/ {N}}}){\mathbf{F}}$, respectively. ${{\mathbf{\Gamma }}_{CP{P_i}}}$ is a diagonal matrix which has an expression of 
\begin{equation}\label{CPP}
\begin{split}
{{\mathbf{\Gamma }}_{CP{P_i}}} = {\text{diag}}\left( {\left\{ {\begin{array}{*{20}{c}}
  {{{\mathbf{e}}^{ - j2\pi {c_1}\left( {{N^2} - 2N\left( {{\ell_i} - n} \right)} \right)}}}&{n < {\ell_i}} \\ 
  1&{n \geqslant {\ell_i}} 
\end{array}} \right.} \right).
\end{split}
\end{equation}

From \eqref{CPP}, we can see that if $N$ is an even number and $2Nc_1$ is an integer, then $\bm{\Gamma}_{CPP_i}=\mathbf{I}$.

\subsection{Received Communication Signal}
After the transmission of AFDM communication signal through a doubly dispersive channel, the received sample of the communication signal at the user side is given by
\begin{equation}
    r_{com}\left[ n \right] = \sum\limits_{l = 1}^L {{\rho_l}{{\text{e}}^{ - j2\pi \frac{f_{d,l}}{N\Delta f}n}}} s\left[ {n - {\ell_l}} \right] + w\left[ n \right],
\end{equation}
where $L$ is the number of paths in the communication channel. ${\rho_l}$, ${f_{d,l}}$, and ${\ell_l}$ are the complex channel gain, Doppler shift, and the normalized delay for the $l$-th path, respectively.

After CPP removal, the signal in the DAF domain can be obtained by demodulation given by
\begin{equation}
    y_{com}\left[ m \right] = \sum\limits_{n = 0}^{N - 1} {r_{com}\left[ n \right]\phi _n^*\left( m \right)}.\label{y[m]}
\end{equation}

This paper primarily addresses sensing advancements. The specifics of communication processing, including channel estimation and symbol detection, have been explored in \cite{AFDM TWC}.  
\section{Proposed Off-grid Parameter Estimation}
\subsection{Problem Formulation}
The received AFDM sensing signal in \eqref{yr} can be rewritten as
\begin{equation}\label{y=Ax}
    {\mathbf{y}} = {\mathbf{A }}\left( {\boldsymbol{\ell},\bm{\nu} } \right)\tilde{\mathbf{ h}} + \tilde{\mathbf{ w}},
\end{equation}
where $\boldsymbol{\ell}=\left[\ell_1,..,\ell_P\right]^T$, $\bm{\nu}=\left[\nu_1,..,\nu_P\right]^T$ and $\tilde{\mathbf{h}}=\left[\tilde{h}_1,..,\tilde{h}_P\right]^T$. ${\mathbf{A }}\left( {\boldsymbol{\ell},\bm{\nu} } \right)=\left[ {{\mathbf{a }}\left( {{\ell_1},{\nu _1}} \right),...,{\mathbf{a }}\left( {{\ell_P},{\nu _P}} \right)} \right] \in {\mathbb{C}^{N \times P}}$ with ${\mathbf{a }}\left( {{\ell_i},{\nu _i}} \right) = {{\mathbf{\Lambda }}_{{c_2}}}{\mathbf{F}}{{\mathbf{\Lambda }}_{{c_1}}}{{\mathbf{\Gamma }}_{CP{P_i}}}{{\mathbf{\Delta }}_{{{{\nu _i}} \mathord{\left/ {\vphantom {{{\nu _i}} N}} \right. \kern-\nulldelimiterspace} N}}}{{\mathbf{\Pi }}^{{\ell_i}}}{\mathbf{\Lambda }}_{{c_1}}^H{{\mathbf{F}}^H}{\mathbf{\Lambda }}_{{c_2}}^H{\mathbf{x}}$.

The unknown normalized delay $\ell_i$ and Doppler $\nu_i$ are the parameters we want to estimate for target sensing. However, we only have information on the maximum delay and Doppler shift in the system. Thus, we can form the estimation problem as a sparse signal recovery problem. Here, we define two types of virtual sampling grids, i.e., the virtual delay grid vector $\bar{\boldsymbol{\ell}}$ and the virtual Doppler grid vector $\bar{\mathbf{k}}$. Let $r_\tau$ and $r_k$ be the resolution of the virtual delay and Doppler grid, respectively.
Since we assume the normalized delay is always an integer, ${{r_\tau }}$ is set to $1$. Define virtual grids ${\bar{\boldsymbol{\ell}}} = {\left[ {{{\bar \ell}_0},{{\bar \ell}_1},...,{{\bar \ell}_{{L_\tau }{K_\nu }-1}}} \right]^T} \in {\mathbb{R}^{{L_\tau }{K_\nu }\times 1}}$ and ${\mathbf{\bar k}} = {\left[ {{{\bar k}_0},{{\bar k}_1},...,{{\bar k}_{{L_\tau }{K_\nu }-1}}} \right]^T} \in {\mathbb{R}^{{L_\tau }{K_\nu }\times 1}}$, where ${L_\tau } = \frac{{{\ell_{\max }} + 1}}{{{r_\tau }}}$ and ${K_\nu } = \frac{{2{\alpha _{\max }}}}{{{r_k}}} + 1$. Let $j = \ell'{L_\tau } + k'$ with $j \in {\left\{0,...,{{L_\tau }{K_\nu }-1}\right\}}$, we have ${{\bar \ell}_j} = \ell'{r_\tau }$ and ${{\bar k}_j} = k'{r_k} - {\alpha _{\max }}$, where $\ell' \in {\left\{0,...,{{L_\tau }-1}\right\}}$ and $k' \in {\left\{0,...,{{K_\nu }-1}\right\}}$. Suppose that ${{\bar k}_{{n_i}}}$ and ${{\bar \ell}_{{n_i}}}$ with $n_i\in \left\{0,1,...,{{L_\tau }{K_\nu }-1}\right\}$ are the nearest virtual grid point to $\nu_i$ and $\ell_i$, respectively. The linear approximation for the column of ${\mathbf{A }}\left( {\boldsymbol{\ell},\bm{\nu} } \right)$ using a first-order Taylor expansion can be expressed as
\begin{equation}\label{Taylor}
    {\mathbf{a}}\left( {{\ell_i},{\nu _i}} \right) \approx {\mathbf{a}}\left( {{{\bar \ell}_{{n_i}}},{{\bar k}_{{n_i}}}} \right) + {\mathbf{b}}\left( {{{\bar \ell}_{{n_i}}},{{\bar k}_{{n_i}}}} \right)\left( {{\nu _i} - {{\bar k}_{{n_i}}}} \right),
\end{equation}
with ${\mathbf{b}}\left( {{{\bar \ell}_{{n_i}}},{{\bar k}_{{n_i}}}} \right) = \frac{{{\mathbf{a}}\left( {{{\bar \ell}_{{n_i}}},{{\bar k}_{{n_i}}}} \right)}}{{\partial {{\bar k}_{{n_i}}}}}$. Define ${\kappa _{{n_i}}} = {\nu _i} - {{\bar k}_{{n_i}}} \in \left[ { - \frac{{{r_k}}}{2},\frac{{{r_k}}}{2}} \right]$ and by incorporating the approximation error into the measurement noise, the observation model in \eqref{y=Ax} can be rewritten as
\begin{equation}\label{y=Hx}
    {\mathbf{y}} = {\mathbf{\Phi }}\left( {\bm{\kappa }} \right){\mathbf{\bar h}} + {\mathbf{\bar w}},
\end{equation}
where the measurement matrix ${\mathbf{\Phi }}\left( {\bm{\kappa }} \right) \in \mathbb{C}^{N \times {L_\tau }{K_\nu }}$ can be expressed as
\begin{equation}
    {\mathbf{\Phi }}\left( {\bm{\kappa }} \right) = {\mathbf{A}} + {\mathbf{B}}{\text{diag}}\left( \bm{\kappa}  \right),
\end{equation}
with ${\mathbf{A}} = \left[ {{\mathbf{a}}\left( {{{\bar \ell}_0},{{\bar k}_0}} \right),{\mathbf{a}}\left( {{{\bar \ell}_1},{{\bar k}_1}} \right),...,{\mathbf{a}}\left( {{{\bar \ell}_{{L_\tau }{K_\nu } - 1}},{{\bar k}_{{L_\tau }{K_\nu } - 1}}} \right)} \right]$, ${\mathbf{B}} = \left[ {{\mathbf{b}}\left( {{{\bar \ell}_0},{{\bar k}_0}} \right),{\mathbf{b}}\left( {{{\bar \ell}_1},{{\bar k}_1}} \right),...,{\mathbf{b}}\left( {{{\bar \ell}_{{L_\tau }{K_\nu } - 1}},{{\bar k}_{{L_\tau }{K_\nu } - 1}}} \right)} \right]$, and $\bm{\kappa}  = {\left[ {{\kappa _0},{\kappa _1},...,{\kappa _{{L_\tau }{K_\nu } - 1}}} \right]^T}$. ${\mathbf{\bar h}}\in {\mathbb{C}^{{L_\tau }{K_\nu }\times 1}}$ is a sparse vector with only $P$ non-zero entries and ${\mathbf{\bar w}} \sim \mathcal{C}\mathcal{N}\left( {{\mathbf{0}},{\beta ^{ - 1}}{{\mathbf{I}}_{{L_\tau }{K_\nu }}}} \right)$ is the unknown noise which contains the noise from the environment, and the error from the approximation of \eqref{Taylor}.

Therefore, the original estimation problem \eqref{y=Ax} is transferred to an off-grid sparse signal recovery problem \eqref{y=Hx}. The unknown vectors $\bar{\mathbf{h}}$ and $\bm{\kappa}$ are jointly sparse with the same support. Their sparsity order is $P$, and the remaining elements in the vectors are all zero, which do not need updating during the iterations in Section \ref{C}. 
\subsection{Sparse Bayesian Formulation}
According to \eqref{y=Hx} and the sparsity of $\mathbf{\bar h}$, we develop the sparse Bayesian model by employing a hierarchical prior that is commonly used in SBL \cite{SBL} \cite{Z. Wei OG}. $\mathbf{\bar h}$ is assumed to follow a zero-mean Gaussian prior distribution which has the expression of
\begin{equation}\label{h_prior}
    p\left( {{\mathbf{\bar h}}|{\bm{\delta }}} \right) = \mathcal{C}\mathcal{N}\left( {{\mathbf{\bar h}};{\mathbf{0}},{\bm{\Delta }}} \right),
\end{equation}
where ${\bm{\Delta }} = {\text{diag}}\left( {\bm{\delta }} \right)$ is the covariance matrix with a diagonal form. ${\bm{\delta }} = \left[ {{\delta _0},{\delta _1},...,{\delta _{{L_\tau }{K_\nu } - 1}}} \right]^T \in {\mathbb{R}^{{L_\tau }{K_\nu }\times 1}}$ is the hyper-parameters used to control the sparsity of $\mathbf{\bar h}$ and is assumed to follow a Gamma distribution as
\begin{equation}
    p\left( {\bm{\delta }} \right) = \prod\limits_{j = 0}^{{L_\tau }{K_\nu }} {\Gamma \left( {{\delta _j};1,b} \right)},
\end{equation}
where $b$ is a small fixed positive root parameter. The value of $\bar{h}_j$ is expected to be $0$ when $\delta_j=0$. And the Gamma distribution is defined as $\Gamma \left( {x;a,b} \right) = \Gamma {\left( a \right)^{ - 1}}{b^a}{x^{a - 1}}{\operatorname{e} ^{ - bx}}$ with $\Gamma \left( a \right) = \int_0^\infty  {{t^{a - 1}}{\operatorname{e} ^{ - t}}} dt$.

Additionally, we assume that the noise vector $\mathbf{\bar w}$ follows a Gaussian distribution of
\begin{equation}
    p\left( {{\mathbf{\bar w}}|\beta } \right) = \mathcal{C}\mathcal{N}\left( {{\mathbf{\bar w}};{\mathbf{0}},{\beta ^{ - 1}}{{\mathbf{I}}_{{L_\tau }{K_\nu }}}} \right),
\end{equation}
where $\beta$ denotes the inverse of the noise variance and is assumed to follow the Gamma distribution as
\begin{equation}
    p\left( \beta  \right) = \Gamma \left( {\beta ;d,e} \right),
\end{equation}
where $d,e>0$ are small fixed root parameters. Furthermore, for the off-grid Doppler variables ${\bm{\kappa }}$, the only information we have is its boundedness. Therefore, we assume that its prior follows a uniform distribution
\begin{equation}
    p\left( {{\kappa _j}} \right) = U\left[ { - \frac{{{r_k}}}{2},\frac{{{r_k}}}{2}} \right],{\text{   }}j \in {\left\{0,1,...,{{L_\tau }{K_\nu }-1}\right\}}.
\end{equation}
\subsection{Off-grid SBL-based Target Parameter Estimation}\label{C}
According to the Bayesian principle as well as the sparse Bayesian model, the optimal estimation for the unknown variables in \eqref{y=Hx} can be obtained through maximizing the posterior probability $p\left( {{\mathbf{\bar h}},{\bm{\delta }},\beta ,{\bm{\kappa }}|{\mathbf{y}}} \right)$. However, due to the non-linearly coupled unknown variables in ${{\bm{\Phi }}\left( {\bm{\kappa }} \right)}$, the posterior probability $p\left( {{\mathbf{\bar h}},{\bm{\delta }},\beta ,{\bm{\kappa }}|{\mathbf{y}}} \right)$ cannot be computed directly. Moreover, the determination of the posterior distribution is significantly complicated by the unspecified hyper-parameters $\bm{\delta}$ and $\beta$ found in the prior distribution and the likelihood function, respectively.

Instead, we decompose the posterior according to Bayesian rule as
\begin{equation}\label{post}
    p\left( {{\mathbf{\bar h}},{\bm{\delta }},\beta ,{\bm{\kappa }}|{\mathbf{y}}} \right) = p\left( {{\mathbf{\bar h}}|{\mathbf{y}},{\bm{\delta }},\beta ,{\bm{\kappa }}} \right)p\left( {{\bm{\delta }},\beta ,{\bm{\kappa }}|{\mathbf{y}}} \right).
\end{equation}

The decomposed expression of \eqref{post} inspires the idea that we can obtain the conditional posterior distribution $p\left( {{\mathbf{\bar h}}|{\mathbf{y}},{\bm{\delta }},\beta ,{\bm{\kappa }}} \right)$ of ${\mathbf{\bar h}}$ given the hyper-parameters $\left({\bm{\delta }},\beta ,{\bm{\kappa }}\right)$, and then update these hyper-parameters by the EM algorithm to maximize the posterior probability $p\left( {{\bm{\delta }},\beta ,{\bm{\kappa }}|{\mathbf{y}}} \right)$. The newly obtained hyper-parameters can in turn re-update the conditional posterior probability. This alternative inference computation will be repeated until convergence.

Given the hyper-parameters $\left({\bm{\delta }},\beta ,{\bm{\kappa }}\right)$, the conditional posterior distribution of $\mathbf{\bar h}$ can be expressed as
\begin{equation}\label{condi_post}
    \begin{split}
        &p\left( {{\mathbf{\bar h}}|{\mathbf{y}};{{\bm{\delta }}^{\left( t \right)}},{\beta ^{\left( t \right)}},{{\bm{\kappa }}^{\left( t \right)}}} \right)\\ &\propto p\left( {{\mathbf{y}}|{\mathbf{\bar h}};{{\bm{\delta }}^{\left( t \right)}},{\beta ^{\left( t \right)}},{{\bm{\kappa }}^{\left( t \right)}}} \right)p\left( {{\mathbf{\bar h}}|{{\bm{\delta }}^{\left( t \right)}}} \right),
    \end{split}
\end{equation}
where the prior distribution $p\left( {{\mathbf{\bar h}}|{{\bm{\delta }}^{\left( t \right)}}} \right)$ is given in \eqref{h_prior} and the likelihood distribution in the $t$-th iteration is given by
\begin{equation}\label{likely}
    p\left( {{\mathbf{y}}|{\mathbf{\bar h}};{{\bm{\delta }}^{\left( t \right)}},{\beta ^{\left( t \right)}},{{\bm{\kappa }}^{\left( t \right)}}} \right) = \mathcal{C}\mathcal{N}\left( {{\mathbf{y}};{\mathbf{\Phi }}\left( {{{\bm{\kappa }}^{\left( t \right)}}} \right){\mathbf{\bar h}},\frac{1}{{{\beta ^{\left( t \right)}}}}{{\mathbf{I}}_{{L_\tau }{K_\nu }}}} \right).
\end{equation}

Combine \eqref{h_prior}, \eqref{condi_post} and \eqref{likely}, the conditional posterior distribution can be express as

\begin{equation}
    p\left( {{\mathbf{\bar h}}|{\mathbf{y}};{{\bm{\delta }}^{\left( t \right)}},{\beta ^{\left( t \right)}},{{\bm{\kappa }}^{\left( t \right)}}} \right) = \mathcal{C}\mathcal{N}\left( {{\mathbf{\bar h}};{{\bm{\mu }}^{\left( t \right)}},{{\bm{\Sigma }}^{\left( t \right)}}} \right),
\end{equation}
with the posterior covariance and the mean of
\begin{equation}\label{sigma}
    {{\mathbf{\Sigma }}^{\left( t \right)}} = {\left( {{\beta ^{\left( t \right)}}{\mathbf{\Phi }}{{\left( {{{\bm{\kappa }}^{\left( t \right)}}} \right)}^H}{\mathbf{\Phi }}\left( {{{\bm{\kappa }}^{\left( t \right)}}} \right) + {{\mathbf{\Delta }}^{ - 1}}} \right)^{ - 1}},
\end{equation}
and
\begin{equation}\label{miu}
    {{\bm{\mu }}^{\left( t \right)}} = {\beta ^{\left( t \right)}}{{\mathbf{\Sigma }}^{\left( t \right)}}{\mathbf{\Phi }}{\left( {{{\bm{\kappa }}^{\left( t \right)}}} \right)^H}{\mathbf{y}}.
\end{equation}
\begin{figure*}
	\setlength{\abovecaptionskip}{-5pt}
	\setlength{\belowcaptionskip}{-10pt}
	\centering
	\begin{minipage}[b]{0.33\linewidth}
		\centering
		\includegraphics[width=2.3in]{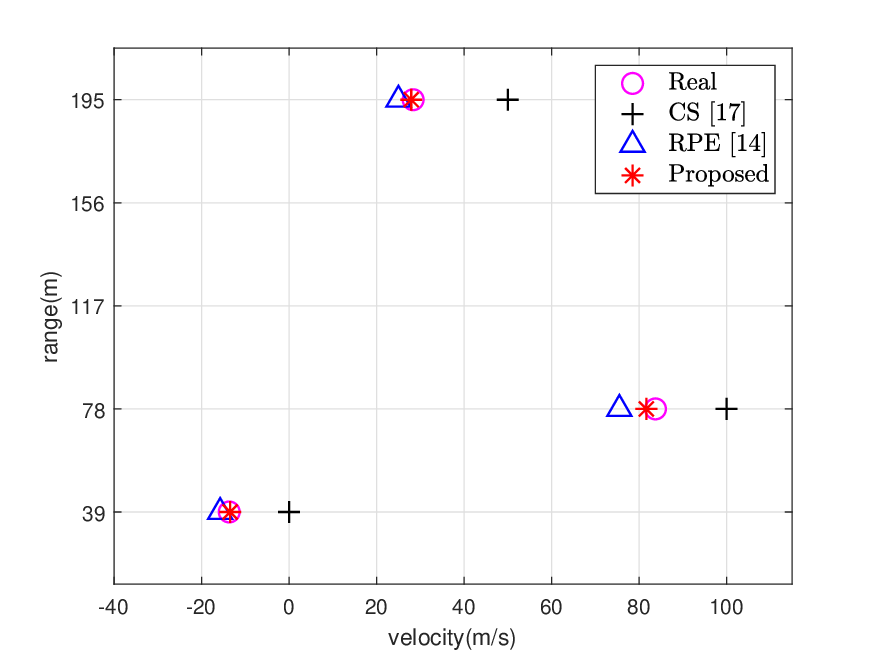}
		\footnotesize{(a) $r_k=0.5$}
		\label{fig3}
	\end{minipage}%
	\begin{minipage}[b]{0.33\linewidth}
		\centering
		\includegraphics[width=2.3in]{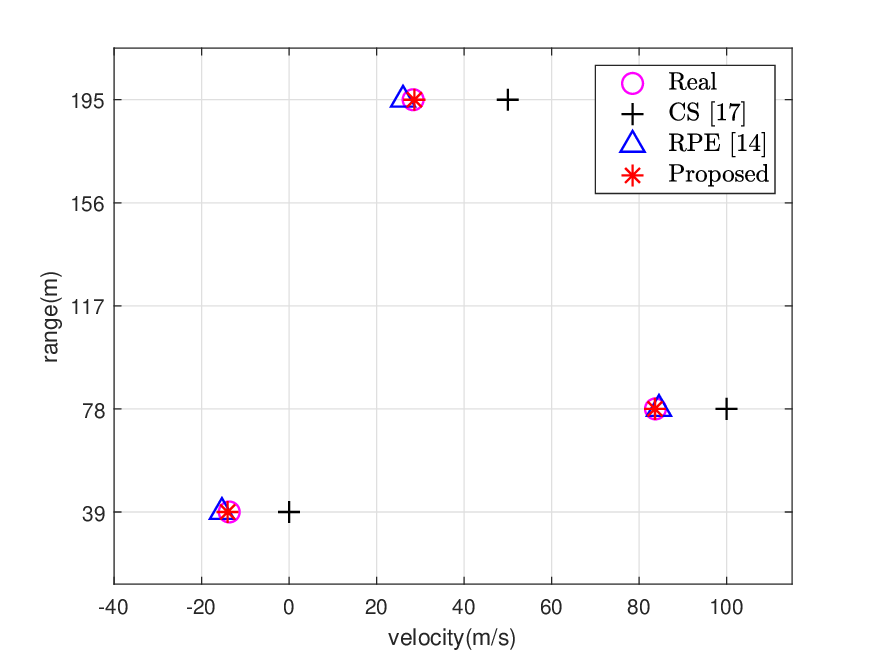}
		\footnotesize{(b) $r_k=0.3$}
		\label{fig4}
	\end{minipage}
	\begin{minipage}[b]{0.33\linewidth}
		\centering
		\includegraphics[width=2.3in]{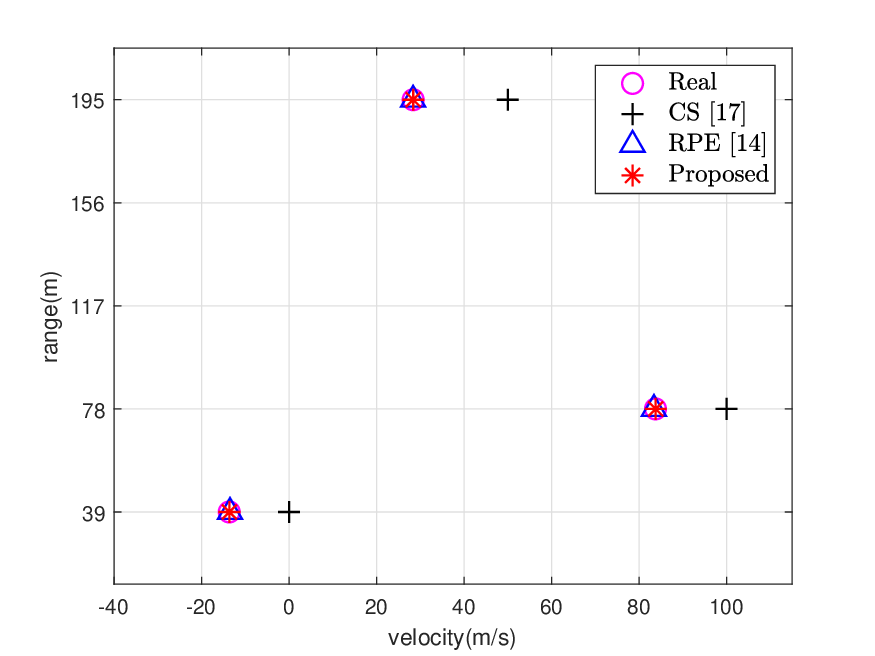}
		\footnotesize{(c) $r_k=0.1$}
		\label{fig5}
	\end{minipage}
 \caption{Range-velocity estimation performance for different virtual resolution $r_k$ at $5$ dB.}
 \label{r-v}
\end{figure*}

Subsequently, given the conditional posterior distribution, the hyper-parameters are updated by maximizing $p\left( {{\bm{\delta }},\beta ,{\bm{\kappa }}|{\mathbf{y}}} \right)$. Since ${\mathbf{y}}$ is independent of these hyper-parameters, it is equivalent to maximize the joint distribution $p\left( {{\bm{\delta }},\beta ,{\bm{\kappa }},{\mathbf{y}}} \right)$. However, the maximization cannot be determined directly. Thus, we focus on the joint distribution of $p\left( {{\mathbf{\bar h}},{\bm{\delta }},\beta ,{\bm{\kappa }},{\mathbf{y}}} \right)$, where ${\mathbf{\bar h}}$ is regarded as a hidden variable via employing the expectation maximization (EM) algorithm for hyper-parameter update. Following the Bayesian rule, the joint distribution can be decomposed as
\begin{equation}
    p\left( {{\mathbf{\bar h}},{\bm{\delta }},\beta ,{\bm{\kappa }},{\mathbf{y}}} \right) = p\left( {{\mathbf{y}}|{\mathbf{\bar h}},\beta ,{\bm{\kappa }}} \right)p\left( {{\mathbf{\bar h}}|{\bm{\delta }}} \right)p\left( {\bm{\delta }} \right)p\left( \beta  \right)p\left( {\bm{\kappa }} \right).
\end{equation}

At the $\left(t+1\right)$-th iteration for the expectation step, we represent the adaptive lower bound of the log joint distribution ${\log \left( {p\left( {{\mathbf{\bar h}},{\bm{\delta }},\beta ,{\bm{\kappa }},{\mathbf{y}}} \right)} \right)}$ as
\begin{equation}
    \begin{split}
        &Q\left( {{\bm{\delta }},\beta ,{\bm{\kappa }}|{{\bm{\delta }}^{\left( t \right)}},{\beta ^{\left( t \right)}},{{\bm{\kappa }}^{\left( t \right)}}} \right) \\
        &= {\mathbb{E}_{{\mathbf{\bar h}}|{\mathbf{y}};{{\bm{\delta }}^{\left( t \right)}},{\beta ^{\left( t \right)}},{{\bm{\kappa }}^{\left( t \right)}}}}\left[ {\log \left( {p\left( {{\mathbf{\bar h}},{\bm{\delta }},\beta ,{\bm{\kappa }},{\mathbf{y}}} \right)} \right)} \right].
    \end{split}
    \label{Q}
\end{equation}

In the maximization step, we can always get the improvement of the joint distribution by updating the hyper-parameters through maximizing \eqref{Q}. The hyper-parameters are updated by

\begin{equation}
    {{\bm{\delta }}^{\left( {t + 1} \right)}} = \mathop {\arg \max }\limits_{\bm{\delta }} Q\left( {{\bm{\delta }}|{{\bm{\delta }}^{\left( t \right)}},{\beta ^{\left( t \right)}},{{\bm{\kappa }}^{\left( t \right)}}} \right),
\end{equation}

\begin{equation}
    {\beta ^{\left( {t + 1} \right)}} = \mathop {\arg \max }\limits_{\beta} Q\left( {\beta |{{\bm{\delta }}^{\left( t \right)}},{\beta ^{\left( t \right)}},{{\bm{\kappa }}^{\left( t \right)}}} \right),
\end{equation}

\begin{equation}
    {{\bm{\kappa }}^{\left( {t + 1} \right)}} = \mathop {\arg \max }\limits_{\bm{\kappa }} Q\left( {{\bm{\kappa }}|{{\bm{\delta }}^{\left( t \right)}},{\beta ^{\left( t \right)}},{{\bm{\kappa }}^{\left( t \right)}}} \right).
\end{equation}

Accordingly, the closed-form updating rule can be given by
\begin{equation}\label{delta}
    \delta _j^{\left( {t + 1} \right)} = \frac{{\sqrt {1 + 4b\left( {{{\left| {{\mathbf{\mu }}_j^{\left( t \right)}} \right|}^2} + {\mathbf{\Sigma }}_{j,j}^{\left( t \right)}} \right)}  - 1}}{{2b}},
\end{equation}

\begin{equation}\label{beta}
    {\beta ^{\left( {t + 1} \right)}} 
    = \frac{{d - 1 + {L_\tau }{K_\nu }}}{{e + \mathbf{T} + \frac{1}{{{\beta ^{\left( t \right)}}}}\sum\limits_{j = 0}^{{L_\tau }{K_\nu }} {\left( {1 - \frac{{{\mathbf{\Sigma }}_{j,j}^{\left( t \right)}}}{{\delta _j^{\left( t \right)}}}} \right)} }}.
\end{equation}
with $\mathbf{T} = {{\left\| {{\mathbf{y}} - {\mathbf{\Phi }}\left( {{{\bm{\kappa }}^{\left( t \right)}}} \right){{\bm{\mu }}^{\left( t \right)}}} \right\|}^2}$. And 
\begin{equation}\label{kappa}
    {\kappa} _j^{\left( {t + 1} \right)} = \frac{{{\eta} _j^{\left( t \right)} - \left\{ {{\bm{\Xi} }_j^{\left( t \right)}} \right\}_{ - j}^T{{\left\{ {{{\bm{\kappa }}^{\left( t \right)}}} \right\}}_{ - j}}}}{{{\Xi} _{j,j}^{\left( t \right)}}},
\end{equation}
where $ \left\{  \cdot  \right\}_{ - j}$ denotes the vector without the $j$-th entry, and
\begin{equation*}
    {{\mathbf{\Xi }}^{\left( t \right)}} = \operatorname{Re} \left\{ {{{\mathbf{B}}^H}{\mathbf{B}} \odot \left( {{{\mathbf{\Sigma }}^{\left( t \right)}} + {{\bm{\mu }}^{\left( t \right)}}{{\bm{\mu }}^{\left( t \right)H}}} \right)} \right\},
\end{equation*}
\begin{equation*}
\begin{split}
    &{{\bm{\eta }}^{\left( t \right)}} = \\
    &\operatorname{Re} \left\{ {{\text{diag}}\left( {{{\bm{\mu }}^{\left( t \right)H}}} \right){{\mathbf{B}}^H}\left( {{\mathbf{y}} - {\mathbf{A}}{{\bm{\mu }}^{\left( t \right)}}} \right) - {\text{diag}}\left( {{{\mathbf{B}}^H}{\mathbf{A}}{{\mathbf{\Sigma }}^{\left( t \right)}}} \right)} \right\}.
\end{split}
\end{equation*}

Due to the fact that $\bm{\kappa}$ is jointly sparse with $\bm{\delta}$, we only need to update the entries of $\bm{\kappa}$ according to the indexes of the largest $P$ values of $\bm{\mu}$ in \eqref{miu}. Thus, $\bm{\kappa }$, $\bm{\eta }$ and $\bm{\Xi }$ can be truncated to $\bar{\bm{\kappa }} \in {\mathbb{R}^{P\times 1}}$, $\bar{\bm{\eta }} \in {\mathbb{R}^{P\times 1}}$ and $\bar{\bm{\Xi }} \in {\mathbb{R}^{P\times P}}$, respectively. The remaining entries of $\bm{\kappa}$ keep to be zero without updating.
Since the value of ${\kappa_j}$ is limited by the prior boundary conditions, the final update rule for ${\kappa_j} \in \left[-\frac{r_k}{2},\frac{r_k}{2}\right]$ also needs to be satisfied.

The iterative process of our proposed off-grid SBL-based target parameters estimation scheme for the AFDM-ISAC system is summarized in \textbf{Algorithm \ref{Algorithm 1}}.
\begin{algorithm}\label{Algorithm 1}
\caption{Proposed Off-grid SBL-based Target Parameter Estimation}\label{algorithm}
\KwIn{${\mathbf{y}}$, ${\ell_{\max }}$, ${\alpha_{\max }}$, ${{\mathbf{\bar k}}}$, ${{\bar{\boldsymbol{\ell}}}}$, parameters $b$, $c$, $d$, the convergence tolerance $\varepsilon$, maximum iteration $I_{\max }$}
\textbf{Initialization}:
$t=0$, ${\beta ^{\left( 0 \right)}} = \frac{{100N}}{{{{\left\| y \right\|}^2}}}$, ${{\bm{\kappa }}^{\left( 0 \right)}} = {\mathbf{0}}$, ${{\bm{\delta }}^{\left( 0 \right)}} = \left| {{{\mathbf{A}}^H}{\mathbf{y}}} \right|$\;
\textbf{Repeat}

Compute ${{\mathbf{\Sigma }}^{\left( t \right)}}$ and ${{\bm{\mu }}^{\left( t \right)}}$ according to \eqref{sigma} and \eqref{miu}, respectively\;
Update the hyper-parameters ${{\bm{\delta }}^{\left( {t + 1} \right)}}$, ${\beta ^{\left( {t + 1} \right)}}$ and ${{\bm{\kappa }}^{\left( {t + 1} \right)}}$ according to \eqref{delta}, \eqref{beta} and \eqref{kappa}, respectively\;
$t\leftarrow t+1$\;
\textbf{until} $ \frac{{\left\| {{{\bm{\delta }}^{\left( {t + 1} \right)}} - {{\bm{\delta }}^{\left( t \right)}}} \right\|_F^2}}{{\left\| {{{\bm{\delta }}^{\left( t \right)}}} \right\|_F^2}} < \varepsilon $ or $t=I_{\max }$\;
Calculate range and velocity according to the indexes of the $P$ largest values of $\bm{\mu}$ \;
\KwOut{ranges and velocities of the targets.}
\end{algorithm}

\section{Numerical Results}

In our system settings, carrier frequency $f_c$ is $90$ GHz, and the chirp subcarrier spacing $\Delta f$ is 30 kHz. The number of chirp subcarriers is set to be $N= 128$ and we employ 16-QAM symbols. The maximum normalized Doppler shift is $\alpha_\text{max}=2$, which corresponds to a maximum velocity of 360 km/h. 
The maximum unambiguous normalized delay $\ell_\text{max}$ is set to be $\ell_\text{max}=10$, and then $N_{cpp}=12$. We set the number of target $P=3$. Without loss of generality, the targets are assumed to have random and distinguished ranges and velocities unless specified. The reflection coefficient $h_i$ is a random variable following a Gaussian distribution.

The root mean square error (RMSE) is adopted for measuring the sensing target parameters estimation performance, which is defined as
\begin{equation}
    {\text{RMSE}} = \frac{1}{P} {\sqrt {{\sum\limits_{i = 1}^P{\left( {\hat {V}_i - {V_i}} \right)}^2}} },
\end{equation}
where $\hat {V}_i$ is the estimated velocity for the $i$-th target.


\begin{figure}
\centering
\includegraphics[width=8.5cm]{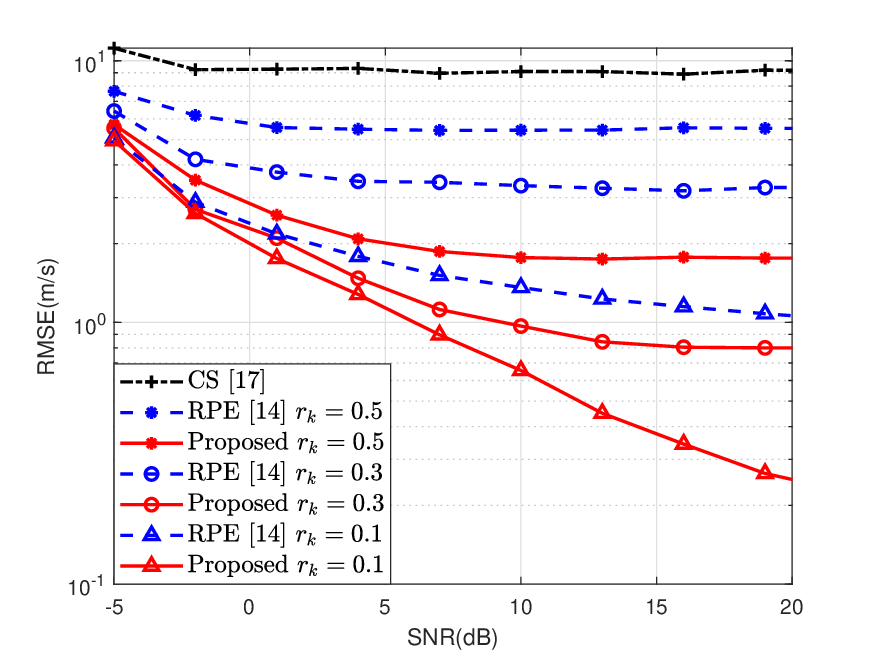}
\caption{RMSE vs SNR.}
\label{SNR}
\end{figure}

We provide the range-velocity estimation performance of our proposed Algorithm \ref{Algorithm 1} with different Doppler virtual grid resolution $r_k$ in Fig.~\ref{r-v}. For comparison, we evaluate the performance of the on-grid method, i.e., the radar parameter estimation (RPE) method in \cite{K. R. R. Ranasinghe} with $\bm{\kappa}=\mathbf{0}$ and the compressed sensing (CS) method in \cite{W. Benzine CS} for AFDM. The SNR is set to be $5$ dB and three targets with the ranges and velocities specific in $\left(39\text{ m},-13.68\text{ m/s}\right)$, $\left(78\text{ m},83.75\text{ m/s}\right)$ and $\left(195\text{ m},28.36\text{ m/s}\right)$ are taken as an example. It can be seen that our proposed off-grid SBL-based target parameter estimation scheme is closer to the real position of the target than the on-grid RPE \cite{K. R. R. Ranasinghe} and CS \cite{W. Benzine CS} at any virtual resolution $r_k$. As $r_k$ decreases (i.e., the resolution is getting better), our method and the on-grid RPE \cite{K. R. R. Ranasinghe} can achieve more accurate sensing estimation. However, the performance of the CS \cite{W. Benzine CS} is not affected by the virtual resolution $r_k$ as it can only handle integer Doppler. Additionally, since we assume that the normalized delays are on the grid in our system, the accuracy of the range estimation is high even in the case of a relatively low SNR. Thus, we only focus on the performance of velocity (i.e., Doppler spread) in the subsequent discussion.

Fig.~\ref{SNR} illustrates the RMSE performance of velocity estimation versus SNR for our proposed off-grid SBL-based target parameter estimation scheme compared with the on-grid RPE \cite{K. R. R. Ranasinghe} the CS \cite{W. Benzine CS}. Since the CS method in \cite{W. Benzine CS} is strictly limited by the sampling rate, only the integer component of normalized Doppler is estimated, leading to the worst performance. The performance improvements of the on-grid RPE \cite{K. R. R. Ranasinghe} and our proposed off-grid SBL-based scheme benefit from the possibility of employing virtual Doppler grids with higher resolution to obtain a more refined estimated codebook. However, with the exact virtual resolution, our method can obtain lower RMSE than the on-grid RPE method. Such performance gain comes from the fact that our proposed off-grid SBL-based scheme can compensate for the loss in resolution by estimating the off-grid components as hyper-parameters, thus enabling more accurate parameter estimation. Moreover, as $r_k$ decreases (i.e., the resolution is getting better), the first-order Taylor expansion in \eqref{Taylor} becomes more accurate, thus enhancing the performance of parameter estimation. However, finer resolution also implies a larger dimension codebook, leading to a higher complexity. Therefore, there is a trade-off between performance and complexity.

\begin{figure}
\centering
\includegraphics[width=8.5cm]{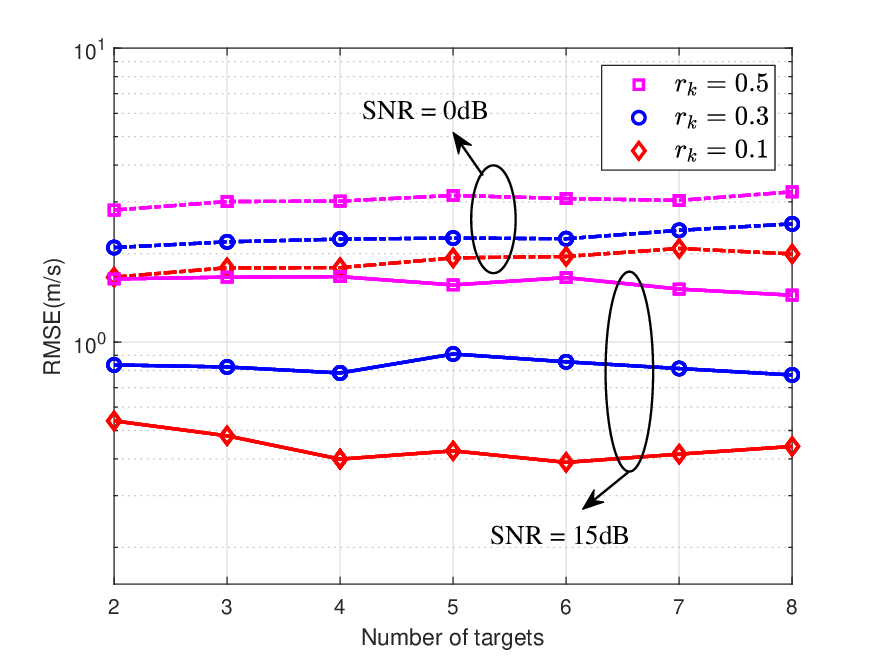}
\caption{RMSE vs number of targets.}
\label{number}
\vspace{-0.9em}
\end{figure}

Finally, we test the RMSE for the velocity estimation performance of our proposed off-grid SBL-based scheme under different numbers of targets in Fig.~\ref{number}. As the number of targets grows, the performance of our proposed scheme can remain essentially stable at both high (15dB) and low (0dB) SNRs, exhibiting strong robustness on target parameter estimation.

\section{Conclusion}
We have proposed a novel target sensing scheme for the high-mobility AFDM-ISAC scenario where the RSU can reuse the reflected communication signal to perform data-aided target parameter estimation. We reformulated the sensing model to a sparse signal recovery problem to exploit the joint sparsity of the channel delay and Doppler. To better estimate Doppler, we set the off-grid part as a hyper-parameter and estimate it separately from the on-grid part via employing the EM algorithm in an SBL framework. Simulation results have demonstrated the estimation accuracy compared with the state-of-the-art method and the strong robustness of our proposed off-grid SBL-based target parameter estimation scheme.

\section*{Acknowledgment}
This work was supported by the RIE2020 Industry Alignment Fund-Industry Collaboration Projects (IAF-ICP) Funding Initiative, as well as cash and in-kind contribution from the industry partner(s).

\vspace{12pt}
\end{document}